\documentclass[letterpaper]{article}
\usepackage{amsmath}
\usepackage{amssymb}

\newcommand{\CN}[1]{#1} 
\newcommand{\OP}[1]{\hat{#1}} 

\newcommand{\PM}{+}
\newcommand{\MP}{-}
\newcommand{\PMf}{}
\newcommand{\MPf}{-}

\newcommand{\tr}{{\mathop{\mathrm{tr}}\nolimits}}
\newcommand{\Tr}{{\mathop{\mathrm{Tr}}\nolimits}}
\newcommand{\Pf}{{\mathop{\mathrm{Pf}}\nolimits}}

\newcommand{\bra}[1]{{\left\langle #1 \right|}}
\newcommand{\ket}[1]{{\left| #1 \right\rangle}}

\begin{document}

\begin{flushright}
UCLA/01/TEP/23\\
hep-th/0110035 
\end{flushright}

\begin{center}

{\Large\bf Non-Commutative Instantons and the Seiberg--Witten Map}\\[5ex]

Per Kraus\footnote{\tt pkraus@physics.ucla.edu} and
Masaki Shigemori\footnote{\tt shige@physics.ucla.edu} 
\\[3ex]

{\it Department of Physics and Astronomy\\
 University of California, Los Angeles\\
 Box 951547, Los Angeles, CA 90095-1547, USA}
\end{center}

\begin{abstract}
We present several results concerning non-commutative instantons and the
Seiberg--Witten map.  Using a simple ansatz we find a large new
class of
instanton solutions in arbitrary even dimensional non-commutative
Yang--Mills theory.  These include the two dimensional ``shift
operator'' solutions and the four dimensional Nekrasov--Schwarz
instantons as special cases.  We also study how the Seiberg--Witten map
acts on these instanton solutions.  The infinitesimal Seiberg--Witten
map is shown to take a very simple form in operator language, and this
result is used to give a commutative description of non-commutative
instantons.  The instanton is found to be singular in commutative
variables.
\end{abstract}

\section{Introduction}

One of the most fruitful applications of non-commutative geometry has
been to the construction of soliton solutions in non-commutative field
theories~\cite{Nekrasov-Schwarz}-\cite{Gross-Nekrasov}.  These solutions
have an elegant operator description, and can often be interpreted as
D-branes in string theory.  It is striking that introducing
non-commutativity---which from one point of view corresponds to adding a
complicated set of higher derivative interactions---can in fact greatly
simplify the construction of soliton solutions.

Our focus here will be on instanton solutions in non-commutative pure
Yang--Mills theory.  Such solutions were originally found by Nekrasov and
Schwarz \cite{Nekrasov-Schwarz} and by Furuuchi \cite{Furuuchi} via a
non-commutative version of the ADHM construction, and have since been
studied by a number of authors.  One interesting fact is the existence
of instantons in non-commutative $U(1)$ gauge theory, since no such
nonsingular and finite action solutions exist in ordinary $U(1)$ gauge
theory.  According to the work of Seiberg and Witten
\cite{Seiberg-Witten}, non-commutative gauge theories are related to
ordinary gauge theories by a change of variables---the Seiberg--Witten
(SW) map.  Starting from non-commutative Yang--Mills theory, one obtains
in this way an ordinary gauge theory with an infinite set of higher
derivative interactions.  It is then natural to ask how the SW map acts
on the known soliton solutions of non-commutative field theories.

For the simplest ``shift operator'' solitons this question was answered
in \cite{K.Hashimoto-Ooguri} using the exact form of the SW map obtained in
\cite{Liu:2000mj,Okawa-Ooguri,Mukhi:2001vx,Liu:2001pk}.  
One typically finds a singular delta
function configuration in the commutative variables.  The corresponding
calculation for the Nekrasov--Schwarz type instanton solutions turns out
to be far more challenging due to the need to compute complicated
symmetrized trace expressions.  We will solve this problem by a more
indirect approach.

In the course of studying this question we have obtained a number of new
results concerning non-commutative instantons and the SW
map.  Non-commutative
solitons are most simply described in the operator formalism, and so it is 
useful to rewrite the infinitesimal SW map in operator form.  The result
turns out to be quite simple.  The resulting SW map states that in mapping
the non-commutativity parameter from $\theta$ to $\theta + \delta \theta$
the operator configuration $X^i$ is mapped to $X^i + \delta X^i$, where
\begin{align}
\delta X^i = {i \over 4} \delta \theta^{kl} \theta_{km} \theta_{ln}
\{ X^m,[X^n,X^i] \}.\label{eq:operator_SW_map_intro}
\end{align}
To put the SW map in this simple form we have used the freedom to
perform unitary transformations, as we discuss in section \ref{sec:NCYM}.

We use
the result (\ref{eq:operator_SW_map_intro}) to follow a non-commutative
instanton solution from finite $\theta=\theta_0$ to $\theta= 0$. 
Based partly on numerical analysis, we solve recursion relations
for the form of the solution in the commutative limit. 
One can freely pass
from the operator formulation to a position space formulation using the
Weyl correspondence, and as with the shift operator solitons we will find
that the instanton solution maps to a singular position space
configuration.  The explicit result for the field strength  is, in
complex coordinates,
\begin{multline}
 F_{\alpha\bar{\beta}}(x;\theta=0)
 = -\frac{i}{\theta_0}
 \Bigg[
 \frac{8(\rho^2-4)}{\rho^2(\rho^4-4\rho^2+8)}\delta^{\bar{\alpha}\beta}\\
 -
 \frac{32(\rho^6-8\rho^4+16\rho^2-16)}{\rho^4(\rho^4-4\rho^2+8)^2}\frac{\bar{z}^{\bar{\alpha}}z^\beta}{\theta_0}
 \Bigg] 
 \qquad (x\neq 0),
\label{FSA}
\end{multline}
where $\rho^2\equiv r^2/\theta_0 = 2\bar{z}z/\theta_0$.
%
An additional singular contribution at the origin gives rise to the
topological charge in the commutative limit.  Thus away from the origin
(\ref{FSA}) gives the commutative description of the Nekrasov--Schwarz
instanton.  The solution has the expected property of shrinking to zero
size as the original noncommutativity parameter $\theta_0$ is taken to
zero.

We have also found a large new class of instanton solutions that
generalize those of \cite{Nekrasov-Schwarz,Furuuchi} in several
directions.  These are found in operator form by starting from the
ansatz 
\begin{align}
 X^\alpha = U f(N) a^\alpha U^\dagger\label{eq:ansatz_intro}
\end{align}
where $U$ is a shift operator.  The equations of motion reduce to a
recursion relation for $f(N)$, which we solve.  This procedure yields
instanton solutions to non-commutative Yang--Mills theory in any even
dimension.  In four dimensions it also generalizes those of
\cite{Nekrasov-Schwarz,Furuuchi} to solutions that are neither self-dual
nor anti-self-dual.  We give explicit examples of these solutions in
dimensions 2 and 4 and evaluate their actions and topological charges in
arbitrary dimensions.

The remainder of this paper is organized as follows.  In section 2  we
first review basic properties of non-commutative gauge theory and then
show that the SW map can be written in a simple covariant form in
operator language.  In section 3, we discuss solitonic solutions in
non-commutative Yang--Mills theory and find new solutions.  In section
4, we will apply the operator SW map to non-commutative solitons and
discuss their commutative description.

\section{Non-commutative gauge theory}\label{sec:NCYM}

%
%

\subsection{Preliminaries and conventions}

Non-commutative  Yang--Mills (NCYM) theory in $D=2d$ dimensional
non-commutative flat space is described by the action
\begin{align}
 S&=-\frac{1}{4g^2} \int {\rm d}^D x\, \tr\,F^{ij} * F_{ij},  \qquad(i,j=1,\dots,2d), \label{eq:ncym_i.t.o._F}\\
 F_{ij} &=\partial_i A_j - \partial_j A_i \MP i (A_i * A_j - A_j * A_i), \label{eq:field_strength}
\end{align}
where the $*$-product is defined by
\begin{align*}
 \CN{f}*\CN{g}(x)=e^{\frac{i}{2}\theta^{ij}\partial_i \partial_j'}\CN{f}(x)\CN{g}(x')\Big|_{x'=x}.
\end{align*}
Under this rule, coordinates become non-commutative:
\begin{align*}
[x^i,x^j]_* \equiv x^i * x^j - x^j * x^i=i\theta^{ij}.
\end{align*}
When necessary, we write the $\theta$ dependence of the $*$-product
explicitly as $*_\theta$.
We consider only the case where the metric is Euclidean: $g_{ij}=\delta_{ij}$.

The action (\ref{eq:ncym_i.t.o._F}) is invariant under the
non-commutative gauge transformation:
\begin{align}
 \delta_\lambda A_i&=\partial_i \lambda \PM i[\lambda,A_i]_*,\\
 \delta_\lambda F_{ij}&= \PMf i[\lambda,F_{ij}]_*,
 \label{eq:nc_gauge_transformation}
\end{align}
where $\lambda(x)$ is an arbitrary infinitesimal parameter.
If we introduce
\begin{align*}
 X^i \equiv x^i \PM \theta^{ij}A_j
\end{align*}
which transforms covariantly under gauge transformation:
\begin{align*}
 \delta_\lambda  X^i = i[\lambda,X^i]_*,
\end{align*}
we can rewrite (\ref{eq:ncym_i.t.o._F}) and (\ref{eq:field_strength}) as
\begin{align}
 S&=-\frac{1}{4g^2} \int {\rm d}^D x\, \tr\, g^{ik} g^{jl}
     \left(i\theta_{im}[X^m,X^n]_*\, \theta_{nj}+\theta_{ij}\right)
     \left(i\theta_{kp}[X^p,X^q]_*\, \theta_{ql}+\theta_{kl}\right), \label{eq:ncym_i.t.o._X}\\
 F_{ij} &= \PMf i\theta_{ik}[X^k,X^l]_*\,\theta_{lj} \PM \theta_{ij},
 \label{eq:field_strength_i.t.o._X}
\end{align}
where $\theta^{ij}\theta_{jk}=\delta^{i}_{k}$.
The equation of motion derived from (\ref{eq:ncym_i.t.o._X}) is
\begin{align}
 g_{ij} [X^i,[X^j,X^k]_*\,]_*=0. \label{eq:eom_i.t.o._X_for_real_coord}
\end{align}

%
%

\bigskip It is often more convenient to work in operator language on
Hilbert space rather than in c-number function language described above.
Define the Weyl transformation $\OP{f}$ \footnote{We denote operators by
hats in this section.}  of a c-number function $f(x)$ by
\begin{align*}
 \OP{f} \equiv
 {\cal W}_{\theta}[\CN{f}(x)]
 \equiv \int {\rm d}^D x\, \frac{{\rm d}^D k}{(2\pi)^D} \,
 \CN{f}(x) e^{-ik\cdot \CN{x}} e^{ik\cdot \OP{x}(\theta)},
\end{align*}
where $\OP{x}^i(\theta)$ are operators satisfying the
operator commutation relation
\begin{align*}
 [\OP{x}^i(\theta),\OP{x}^j(\theta)]
\equiv
 \OP{x}^i(\theta) \OP{x}^j(\theta) -\OP{x}^j(\theta) \OP{x}^i(\theta)
 =i\theta^{ij}.
\end{align*}
We will often drop the argument $(\theta)$ when it is clear.
By the isomorphism
\begin{align*}
 {\cal W}_{\theta}[(f*_\theta g)(x)] = {\cal W}_{\theta}[f(x)] \, {\cal W}_{\theta}[g(x)],
\end{align*}
we can work in c-number function language or equivalently in operator
language. The inverse of the Weyl transformation is
\begin{align*}
 \CN{f}(x) = {\cal W}_{\theta}^{-1}[\OP{f}]
 &\equiv \int \frac{{\rm d}^D k}{(2\pi)^D} \,
 \Pf(2\pi\theta)\, \Tr\,[\OP{f} e^{-ik\cdot \OP{x}(\theta)}] e^{ik\cdot \CN{x}}.
\end{align*}

\subsection{The Seiberg--Witten map in operator language}

It is known \cite{Seiberg-Witten} that a non-commutative gauge theory
with non-commutativity parameter $\theta$ can be equivalently described
by another non-commutative gauge theory with different non-commutativity
parameter $\theta+\delta\theta$.  The relation among the fields in the
two descriptions is given by the so-called Seiberg--Witten (SW) map
\begin{align}
 \CN{A}_i(\theta+\delta\theta)
 &= \CN{A}_i(\theta) + \delta \CN{A}_i(\theta)
 = \CN{A}_i(\theta)
 -\frac{1}{4}\delta\theta^{kl} \{\CN{A}_k(\theta),\partial_l
 \CN{A}_i(\theta)+\CN{F}_{li}(\theta)\}_{*}
 + {\cal O}(\delta\theta^2),\nonumber\\
 \CN{\lambda}(\theta+\delta\theta)
 &= \CN{\lambda}(\theta) + \delta \CN{\lambda}(\theta)
 = \frac{1}{4} \delta\theta^{kl}\{\partial_k
 \CN{\lambda}(\theta),\CN{A}_l(\theta)\}_{*}
 + {\cal O}(\delta\theta^2), \label{eq:original_SW_map_for_c-num_function}
\end{align}
where 
$\{f,g\}_{*}\equiv f*g+g*f$.  The products on the right hand side are
understood as $*_\theta$ products, while the fields on the left hand
side are to be used with $*_{\theta+\delta\theta}$.  We often display
the non-commutativity parameter explicitly as, {\it e.g.},
$A_i(\theta)$, in order to indicate which non-commutative gauge theory
is being referred to.  Note that the new fields
$A_i(\theta+\delta\theta)$, $F_{ij}(\theta+\delta\theta)$ do not
generally satisfy the equation of motion satisfied by the old fields
$A_i(\theta)$, $F_{ij}(\theta)$, since the lagrangian also gets 
transformed.

This map was originally \cite{Seiberg-Witten} (see also \cite{Brace-Cerchiai-Zumino})
derived by the condition that gauge
transformations in the two descriptions are equivalent in the sense that
\begin{align}
 A(A(\theta);\theta+\delta\theta) 
 + \delta_{\lambda(\theta+\delta\theta)} A(A(\theta);\theta+\delta\theta) 
 =
 A(A(\theta)+\delta_{\lambda(\theta)}A(\theta);\theta+\delta\theta).
 \label{eq:SW's_original_logic}
\end{align}
The gauge parameter $\lambda(\theta+\delta\theta)$ is allowed to depend
not only on $\lambda(\theta)$ but also on the gauge field $A(\theta)$.
It is known that the condition (\ref{eq:SW's_original_logic}) does not
determine the map uniquely --- there are infinitely many different
solutions.  However, different maps are related by gauge transformations
and field redefinitions \cite{Asakawa-Kishimoto}.

\bigskip Now let us consider translating the SW map above into
operator language.  The SW map of the
covariant position $\CN{X}^i(\theta)$ in c-number function
language is
\begin{align}
 \CN{X}^i(\theta+\delta\theta)
 &= \CN{X}^i(\theta) + \delta\theta^{ij} \CN{A}_j(\theta)
 + \theta^{ij} \delta \CN{A}_j(\theta) + {\cal O}(\delta\theta^2) \nonumber\\
 &= \CN{X}^i(\theta) + \frac{i}{4}
 \delta\theta^{jk} \theta_{jm} \theta_{kn}
 \{ \CN{X}^m(\theta)-x^m, [\CN{X}^n(\theta)-x^n,
 \CN{X}^i(\theta)]_*\}_* + {\cal O}(\delta\theta^2)\nonumber\\
 &\equiv \CN{X}^i(\theta) + \delta \CN{X}^i(\theta).
 \label{eq:SW_map_for_X_function}
\end{align}
The corresponding operator is
\begin{align}
 \OP{X}^i(\theta+\delta\theta)
 &= {\cal W}_{\theta+\delta\theta}[\CN{X}^i(\theta+\delta\theta)]
 = {\cal W}_{\theta+\delta\theta}[\CN{X}^i(\theta)+\delta\CN{X}^i(\theta)]\nonumber\\
 &= {\cal W}_{\theta+\delta\theta}[X^i(\theta)]
 + \delta \OP{X}^i(\theta)  + {\cal O}(\delta\theta^2), \label{eq:SW_map_for_X_op_intermed}
\end{align}
where $\delta \OP{X}^i(\theta)$ is obtained by replacing $\CN{X}$
functions in $\delta \CN{X}^i(\theta)$ with $\OP{X}$
operators.  The $\OP{x}^i(\theta+\delta\theta)$ operators satisfying
\begin{align*}
 [\OP{x}^i(\theta+\delta\theta),\OP{x}^j(\theta+\delta\theta)]=i(\theta+\delta\theta)^{ij}
\end{align*}
which are necessary for defining ${\cal W}_{\theta+\delta\theta}$ are
most simply constructed, in terms of $\OP{x}^i(\theta)$ operators, as
\begin{align*}
 \OP{x}^i(\theta+\delta\theta)
 = \OP{x}^i(\theta)+ \frac{1}{2}  \delta \theta^{ij}  \theta_{jk} \OP{x}^k(\theta)
 \equiv \OP{x}^i(\theta)+\delta \OP{x}^i(\theta).
\end{align*}
With this choice,
\begin{align*}
 {\cal W}_{\theta+\delta\theta}[\CN{X}^i(\theta)]
 &= \int {\rm d}^D x\, \frac{{\rm d}^D k}{(2\pi)^D} \,
 \CN{X}^i(\theta)\, e^{-ik\cdot x} e^{ik\cdot \OP{x}(\theta+\delta\theta)}\nonumber\\
 &= \int {\rm d}^D x\, \frac{{\rm d}^D k}{(2\pi)^D} \,
 \CN{X}^i(\theta)\, e^{-ik\cdot x} 
 (e^{ik\cdot \OP{x}(\theta)}+ \delta \OP{x}^j(\theta) [\hat{\partial}_j,e^{ik\cdot
 \OP{x}(\theta)}])  \nonumber\\
 &= {\cal W}_{\theta}[X^i(\theta)]
 + \delta \OP{x}^j(\theta) [\OP{\partial}_j,{\cal W}_{\theta}[X^i(\theta)]]\nonumber\\
 &= \OP{X}^i(\theta)
 + \frac{1}{2} \{\delta \OP{x}^j(\theta), [\hat{\partial}_j,\OP{X}^i(\theta)] \},
\end{align*}
where $\OP{\partial}_i\equiv -i \theta_{ij} \OP{x}^j$.  This cancels
some terms in (\ref{eq:SW_map_for_X_op_intermed}) to give\footnote{
Polychronakos \cite{Polychronakos} pointed out that the operator SW map
presented in the older version of this paper was not hermitian.
Eq.~(\ref{eq:SW_map_of_X_operator_with_`gauge'_left}) is the hermitian
form (which was presented in the footnote of the older version).}
\begin{align}
 \delta \OP{X}^i(\theta)
 &= \frac{i}{4} \delta\theta^{kl}\theta_{km}\theta_{ln}
 \{ \OP{X}^m(\theta) , [\OP{X}^n(\theta),\OP{X}^i(\theta)] \}
 + i [ \OP{g},\OP{X}^i(\theta)],
 \label{eq:SW_map_of_X_operator_with_`gauge'_left}
\end{align}
where
\begin{align}
 \OP{g} &\equiv -\frac{1}{4} \delta\theta^{kl}\theta_{km}\theta_{ln}
 \{\OP{x}^m(\theta),\OP{X}^n(\theta)\}
 \label{eq:`gauge'_operator}.
\end{align}
As we show in appendix
\ref{sec:SW_and_SW'_different_by_unitary_transf}, the solution to Eq.\
(\ref{eq:SW_map_of_X_operator_with_`gauge'_left}) and the solution to
the same equation without the last term $i[\OP{g},\OP{X}^i(\theta)]$
are related by unitary transformation.  Namely, if we denote the two
solutions as $\OP{X}_g(\theta)$ and $\OP{X}_0(\theta)$,
respectively, we can always find a unitary operator $\OP{u}(\theta)$
satisfying $\OP{X}_g(\theta)=\OP{u}(\theta) \OP{X}_0(\theta)
\OP{u}(\theta)^\dagger$.  Since unitary transformation is always a symmetry
of non-commutative gauge theory, we can eliminate the last term in
(\ref{eq:SW_map_of_X_operator_with_`gauge'_left}) by performing suitable
unitary transformation at each $\theta$.  
Note that the
unitary transformation $\OP{u}(\theta)$ does not affect the crucial condition
(\ref{eq:SW's_original_logic}), from which the SW map
(\ref{eq:SW_map_for_X_function}) was derived.  In general, the unitary
transformation $\OP{u}(\theta)$ does not correspond to a local gauge
transformation, since the operator $\OP{g}$ is not in general
compact\footnote{An operator $\OP{\cal O}$ on a Hilbert space $\cal H$
is called compact if for any bounded sequence $\{\ket{\psi_i}\}$
($\ket{\psi_i}\in{\cal H}$), the sequence $\{\OP{\cal O}\ket{\psi_i}\}$
contains a convergent subsequence.  For example,
$\OP{P}_L=\ket{0}\bra{0}+\ket{1}\bra{1}+\cdots+\ket{L-1}\bra{L-1}$ is
compact while $\OP{S}=\sum_{n=0}^{\infty}\ket{n}\bra{n+1}$ is noncompact
since the image of the sequence $\{\ket{0},\ket{1},\ket{2},\dots\}$
contains no convergent subsequence.} \cite{Harvey}.

With the above unitary transformation understood, the SW
map becomes
\begin{align}
 \delta \OP{X}^i
 &=
\frac{i}{4} \delta\theta^{kl}\theta_{km}\theta_{ln}
 \{ \OP{X}^m, [\OP{X}^n,\OP{X}^i]\} ,\quad
%
 \delta \OP{\lambda}=0.\label{eq:operator_SW_map}
\end{align}
Note that only
covariant quantities appear on the right hand side, which implies that
the gauge transformation operators at $\theta$ and $\theta+\delta\theta$
are the same.  

In fact, we could have derived the operator SW map above
directly from the condition (\ref{eq:SW's_original_logic}) translated
into operator language:
\begin{enumerate}
 \item  Gauge transformation of $\OP{X}(\theta)$ should lead to a gauge
	transformation of $\OP{X}(\theta+\delta \theta)$.  This is obviously
	satisfied by taking the map to depend only on the covariant quantity
	$\OP{X}(\theta)$ (and consequently
	$\OP{\lambda}(\theta+\delta\theta)=\OP{\lambda}(\theta)$).
 \item If we insert $\OP{X}^i(\theta) = \OP{x}^i(\theta)$ then the
	map should yield $\OP{x}^i(\theta + \delta \theta)$
	satisfying $[\OP{x}^i(\theta + \delta \theta),
	\OP{x}^j(\theta+ \delta \theta)] = i (\theta+ \delta
	\theta)^{ij}$.  This is equivalent to the statement that
	$\OP{A}=0$ should be preserved under the map.
\end{enumerate}
Eq.\ (\ref{eq:operator_SW_map}) gives
explicitly one possible solutions to the above conditions.  One can
easily write down other solutions, but they all differ just by local or
global gauge transformations and field redefinitions
\cite{Asakawa-Kishimoto}.

\bigskip%
The simplest example of the map (\ref{eq:operator_SW_map}) is
\begin{align*}
 \OP{X}^i(\theta)=a^i{}_j(\theta) \OP{x}^i(\theta),
\end{align*}
from which one obtains
\begin{align*}
 2 \delta a= a\theta \delta\theta^{-1}-a\theta a^{\rm T}
 \delta\theta^{-1} a.
\end{align*}
For $D=2$ ($d=1$), this can be solved explicitly to give
\begin{align*}
 a^i{}_j(\theta)=
 \left[ |a_0| - (|a_0|-1)\frac{\theta}{\theta_0} \right]^{-\frac{1}{2}}
 a^i{}_j(\theta_0),
\end{align*}
where $\theta\equiv\theta^{12}$ and
$|a_0|\equiv\det[a^i{}_j(\theta_0)]$. The corresponding field strength
is
\begin{align*}
 F_{12}(\theta)=\frac{|a_0|-1}{|a_0|\theta_0-(|a_0|-1)\theta}
 = \frac{F_{12}(\theta_0)}{1-(\theta-\theta_0)F_{12}(\theta_0)},
\end{align*}
which is singular at $\theta$ if
$F_{12}(\theta_0)=\frac{1}{\theta-\theta_0}$.  This is consistent
with the result of \cite{Seiberg-Witten} that a  commutative description
($\theta=0$) is impossible if $F_{ij}(\theta_0)=(\theta_0^{-1})_{ij}$.

\section{Solitonic solutions in pure NCYM}

Pure NCYM theory has solitonic solutions
\cite{Nekrasov-Schwarz,Polychronakos:2000zm,HKL,Gross-Nekrasov}, some of which have
counterparts in commutative theory and some of which do not.  In this
section, we first review shift operator solitons briefly, and present a
new family of solitonic solutions.  

\subsection{Shift operator solitons}

Shift operator solitons \cite{HKL,Gross-Nekrasov} are
obtained from an arbitrary field $X^i_0$  satisfying the equations of
motion  by applying an``almost gauge transformation''\footnote{In this section,
hats on operators are omitted since we will work only in operator
language.}
\begin{align}
 X^i&=U^\dagger X^i_0 U,\nonumber\\
 UU^\dagger&=1,\quad U^\dagger U=1-P,
 \label{eq:almost_gauge_transf}
\end{align}
where $P$ is a projection operator of a finite rank\footnote{Actually
there is more freedom to add finite dimensional matrices to
(\ref{eq:almost_gauge_transf}) corresponding to the position of the
solitons.  However we do not consider this generalization in this paper
for simplicity. See \cite{Gross-Nekrasov}.
\label{footnote:position_moduli_of_shift_operator_solitons}}.  The
$X^i$ automatically satisfy the equations of motion  because
of the property $UU^\dagger=1$.  Note that the gauge group is
unspecified; the $X_0^i$, $U$, and $X^i$ operators do not have to be $U(1)$,
{\it i.e.\/}, they can be matrices whose entries are operators.  The
simplest example of $X^i_0$ is the vacuum, for which
\begin{align}
 X^i=U^\dagger x^i U
 \label{eq:shift_operator_solitons}
\end{align}
and the field strength is
\begin{align}
 F_{ij}=\theta_{ij} P. 
 \label{eq:field_strength_of_shift_operator_solitons}
\end{align}

Essentially, $U$ is a shift operator which maps one to one the
subspace $(1-P) {\cal H}$ to the whole space ${\cal H}$, while
annihilating the subspace $P {\cal H}$.  For example, in the
$D=2$ ($d=1$) case, 
\begin{align}
 U=\sum_{n^1=0}^\infty \ket{n^1} \bra{n^1+l} \label{eq:2D_shift_operator}
\end{align}
satisfies
\begin{align*}
 U U^\dagger=1,\quad
 U^\dagger U=1-\sum_{n^1=0}^{l-1} \ket{n^1} \bra{n^1}.
\end{align*}
In higher dimensional cases, the non-commutative ABS construction
\cite{Harvey-Moore,HKL} can be used to construct a $U$ operator 
if the gauge group contains a $SO(2d)$ subgroup.

\subsection{NS-type instantons}

Let us consider the case where $\theta^{ij}$ takes the form
\begin{eqnarray}
 \theta^{ij}=
 \left(
  \begin{array}{ccc}
   \begin{array}{cc} 0 & \theta \\ -\theta & 0 \end{array}  & & \text{\Large O}\\
   & \ddots & \\
   \text{\Large O} && \begin{array}{cc} 0 & \theta \\ -\theta & 0 \end{array} 
  \end{array}
 \right),
 \qquad \theta>0,
 \label{eq:homogeneous_non-commutativity}
\end{eqnarray}
when skew-diagonalized.
Take complex coordinates 
\begin{align}
 z^a&=(x^{2a-1}+ix^{2a})/\sqrt{2},\quad
 \bar{z}^{\bar{a}}=(x^{2a-1}-ix^{2a})/\sqrt{2},
 \quad a=1,2,\dots,d,
 \label{eq:complex_coords}
\end{align}
so that
\begin{align}
 [z^\alpha,\bar{z}^{\bar{\beta}}] =\theta \delta^{\alpha\bar{\beta}},
 \qquad \alpha=1,\dots,d;~ \bar{\beta}=\bar{1},\dots,\bar{d},
 \label{eq:zz_algebra}
\end{align}
where 
$\delta^{1\bar{1}}=1$, $\delta^{1\bar{2}}=0$, etc.
The equation of motion (\ref{eq:eom_i.t.o._X_for_real_coord}) can be
written in complex coordinates as
\begin{align}
 [X^\alpha,[X^{\bar{\alpha}},X^\beta]]
 +[X^{\bar{\alpha}},[X^\alpha,X^\beta]]=0,\label{eq:eom_i.t.o._X_for_complex_coord}
\end{align}
where summation over identical barred and unbarred Greek letters is implied.
The algebra (\ref{eq:zz_algebra}) can be realized on the Hilbert space
${\cal H} =\{\ket{n^1,\dots, n^d};~n^1,\dots,n^d=0,1,2,\dots.\}$ by
\begin{align*}
 z^{\alpha}&=\sqrt{\theta}\, a^{\alpha}, \quad
 \bar{z}^{\bar{\alpha}}=\sqrt{\theta}\, a^{\bar{\alpha}},\\
 [a^\alpha,a^{\bar{\beta}}]&=\delta^{\alpha\beta},
 \quad
 [a^\alpha,a^\beta]=[a^{\bar{\alpha}},a^{\bar{\beta}}]=0.
\end{align*}
For simplicity, we take $\theta=1$ henceforth in this section.  Explicit
$\theta$ dependence can be recovered on dimensional grounds.

\bigskip Now let us look for the solution to the equation of motion
(\ref{eq:eom_i.t.o._X_for_complex_coord}), taking an
ansatz\footnote{Some solutions of this form were obtained independently
in \cite{Nekrasov_Le_Houches}.}
\begin{align}
 \begin{array}{l}
 X^\alpha =
 U_l f(N) \, a^\alpha U^\dagger_l \equiv  U_l X_0^\alpha U^\dagger_l,\\
 X^{\bar{\alpha}} = (X_0^\alpha)^\dagger =
 U_l a^{\bar{\alpha}} f(N) U^\dagger_l \equiv  U_l X_0^{\bar{\alpha}} U^\dagger_l,
 \end{array}\label{eq:ansatz}
\end{align}
where
\begin{align}
 U_l U^\dagger_l &=1, 
 \quad  U^\dagger_l U_l = 1-P_l=\theta(N\ge l),
 \quad l=1,2,3,\dots, \label{eq:U+U=projection_operator}\\
 P_l &\equiv \sum_{|n|\le l-1}
 \ket{\{n\}}\bra{\{n\}},\quad 
 N=a^{\bar{\alpha}} a^{\alpha},\quad 
 f(N)^\dagger=f(N).
\end{align}
Here $\{n\}$ is a shorthand notation for $(n^1,\dots,n^d)$, and $|n|\equiv
n^1+\cdots +n^d$.  The function $\theta(P)$ is 1 if the proposition $P$
is true and 0 if $P$ is false. 
For the action (\ref{eq:ncym_i.t.o._X}) to be
finite, we require that
\begin{align}
 f(N) &\to 1  \quad (N \to \infty). \label{eq:f(N)->1}
\end{align}
{}From the shifting property of $U_l$, we can set without
loss of generality
\begin{align}
  f(0)&=f(1)=\cdots=f(l-1)=0, \quad
 \text{or}\quad
 f(N)P_l=P_l f(N)=0,
 \label{eq:f(0)=...=f(l)=0}
\end{align}
because these are projected out and do not enter $X^{\alpha}$,

\bigskip
For this ansatz, the left hand side of the
equation of motion (\ref{eq:eom_i.t.o._X_for_complex_coord}) is
\begin{align}
 [X^\alpha&,[X^{\bar{\alpha}},X^\beta]]
 +[X^{\bar{\alpha}},[X^\alpha,X^\beta]]\nonumber\\
 &=
 U_l\Big[
 f(N)^3 \theta(N+1\ge l)\theta(N\ge l)a^\alpha 
  a^{\bar{\alpha}} a^\beta \nonumber\\
 &\qquad\qquad - f(N) f(N+1)^2 \theta(N+1\ge l)\theta(N+2\ge l) 
  a^\alpha a^\beta a^{\bar{\alpha}} \nonumber\\
 &\qquad\qquad - f(N) f(N-1)^2 \theta(N-1\ge l) \theta(N\ge l)
   a^{\bar{\alpha}} a^\beta a^\alpha \nonumber\\
 &\qquad\qquad + f(N)^3 \theta(N+1\ge l)\theta(N\ge l)
  a^\beta a^{\bar{\alpha}} a^\alpha
 \Big] U^\dagger_l \nonumber \\
 &=
 U_l f(N) 
 \Big[ (2N+d+1)f(N)^2 \nonumber\\
 &\qquad\qquad -(N+d+1) f(N+1)^2 - N f(N-1)^2 
 \Big] a^\beta U^\dagger_l. \label{eq:LHS_of_eom}
\end{align}
Here we used  relations such as
\begin{align*}
 a^\alpha f(N) = f(N+1) a^\alpha,
 \quad
 a^\alpha \theta(N\ge l) = \theta(N+1\ge l) a^\alpha
\end{align*}
as well as (\ref{eq:f(0)=...=f(l)=0}).  Because there are only states
with $N\ge l$ between $U_l$ and $U^\dagger_l$, we observe that the
equation of motion is satisfied if
\begin{align}
 (2N+d+1)f(N)^2 -(N+d+1) f(N+1)^2 - &N f(N-1)^2 =0, \nonumber\\
 & N=l,l+1,l+2,\dots. \label{eq:recursion_equation}
\end{align}
Solving this recursion equation under the initial condition
\begin{align*}
 \begin{array}{ll}
 f(l)=f(l+1)=\cdots=f(L-1)=0,\quad
 f(L)\neq 0,
 \quad l \le L
\end{array}
\end{align*}
along with (\ref{eq:f(N)->1}),
we obtain
\begin{align}
 f(N)&=
 \sqrt{1-\frac{L(L+1)(L+2)\dots(L+d-1)}{(N+1)(N+2)\dots(N+d)}}
 \;\theta(N\ge L)
 \equiv f_L(N). \label{eq:the_solution}
\end{align}
Even if we take into account the $f(N)$ factor in (\ref{eq:LHS_of_eom}),
we end up with the same result (\ref{eq:the_solution}).  In addition, we
could start with more general $U$ and $U^\dagger$ operators which satisfy
\begin{align*}
 U^\dagger U&=\theta(N\not\in{\cal P}),
\end{align*}
instead of (\ref{eq:U+U=projection_operator}), where ${\cal P}$ is a
finite subset of ${\bf N}=\{0,1,2,\dots\}$.  However, this again leads
to (\ref{eq:the_solution}), with $L$ larger than any elements of ${\cal
P}$.  We will refer to the new class of solutions (\ref{eq:ansatz}) as
Nekrasov--Schwarz (NS)-type instantons, because as we will see later
these include the Nekrasov--Schwarz instanton
\cite{Nekrasov-Schwarz,Furuuchi} as a special case.

\bigskip
The field strength of the NS-type instanton is
\begin{align}
 F_{\alpha\beta}&=F_{\bar{\alpha}\bar{\beta}}=0,\quad
 F_{\alpha\bar{\beta}}
 =U_l (F_0)_{\alpha\bar{\beta}} U^\dagger_l, \label{eq:field_strength_of_ansatz-1}
\end{align}
where
\begin{align}
 (F_0)_{\alpha\bar{\beta}}
 &=
 \MPf i \Big[
 \delta^{\bar{\alpha}\beta}\theta(l\le N \le L-1)\nonumber\\
 &\qquad\qquad + \frac{L(L+1)\dots(L+d-1)}{N(N+1)\dots(N+d)}
 \theta(N\ge L)(N\delta^{\bar{\alpha}\beta}-d\,a^{\bar{\alpha}}
 a^\beta)
 \Big].  \label{eq:field_strength_of_ansatz-2}
\end{align}
The action and the topological charge are computed in
appendix \ref{section:action_and_Chern_chars}:
\begin{align}
 S
 &=(2\pi)^d \frac{d}{2}
 [d\, {\cal N}_d(L)-{\cal N}_d(l)],\label{eq:value_of_action}\\
 Q&\equiv \frac{-1}{(\MPf 2\pi)^d d!} \int \wedge^d F
 = \begin{cases}
     {\cal N}_d(l) & (d\ge 2), \\
    -(L-l) & (d=1).
    \end{cases} \label{eq:value_of_topological_charge}
\end{align}
where ${\cal N}_d(L)\equiv\tfrac{L(L+1)\cdots(L+d-1)}{d!}$ is the number
of states with $N\le L-1$.  The topological charge is equal to the
number of states removed by the $U_l$ operator, except for the $D=2$
($d=1$) case.  Note that in each topological class, the action is
minimized when $L=l$ (remember that $L \ge l$).

\bigskip
\subsection{Examples of NS-type instantons}

\noindent
{\bf $\boldsymbol{D=2}$ ($\boldsymbol{d=1}$)}

\noindent
{}From (\ref{eq:the_solution}),
\begin{align}
 f_L(N)=
  \sqrt{\frac{N-L+1}{N+1}} \; \theta(N\ge L).
 \label{eq:the_solution:D=2}
\end{align}
It follows from (\ref{eq:2D_shift_operator}) and
(\ref{eq:field_strength_of_ansatz-1}) that
\begin{align}
 X^1&=U_l f_L(N) a^1 U^\dagger_l = U^\dagger_{L-l} a^1 U_{L-l},\label{eq:we_end_up_with_HKL_solution_in_D=2_case}\\
 (F_0)_{1\bar{1}}&= i \theta(l\le N\le L-1),\quad
 F_{1\bar{1}}= i P_{L-l}.
\end{align}
Therefore in this case the shift operator soliton
(\ref{eq:shift_operator_solitons}) and the NS-type instanton are the
same.  This is consistent with the result \cite{Gross-Nekrasov} that the
most general soliton solution in 2-dimensional pure NCYM is of this form
up to translation.

\bigskip
\noindent
{\bf $\boldsymbol{D=4}$ ($\boldsymbol{d=2}$)}

\noindent
{}From (\ref{eq:the_solution}),
\begin{align}
 f_L(N)=
  \sqrt{\frac{(N-L+1)(N+L+2)}{(N+1)(N+2)}} \; \theta(N\ge L).
 \label{eq:the_solution:D=4}
\end{align}
Note that we can rewrite the solution in a rather suggestive way:
\begin{align*}
 X^{\alpha}&=U_l \xi^{-1} a^{\alpha} \xi U^\dagger_l,\quad
 X^{\bar{\alpha}}=U_l \xi a^{\bar{\alpha}} \xi^{-1}
 U^\dagger_l,\\
 \xi&=
 \sqrt{\frac{(N+2)(N+3)\dots(N+L+1)}{N(N-1)\dots(N-L+1)}} \; \theta(N\ge L)\\
 &=\sqrt{
  \frac{
   a^{\alpha_1}\dots a^{\alpha_L} a^{\bar{\alpha}_L} \dots a^{\bar{\alpha}_1}
  }{
   a^{\bar{\alpha}_1} \dots a^{\bar{\alpha}_L} a^{\alpha_L} \dots a^{\alpha_1}
  } 
 }\; \theta(N\ge L),
\end{align*}
where the inverse $\xi^{-1}$ is defined only on $(1-P_L){\cal H}$.

This solution gives anti-self-dual field strength if and only if
$L=l$.  Although one can see this from the explicit form of the field
strength (\ref{eq:field_strength_of_ansatz-1}), let us show it in a
different way.  In complex coordinates, the anti-self-duality condition
can be written as
\begin{align}
 [X^1,X^2]=0, \quad [X^1,X^{\bar{1}}]+[X^2,X^{\bar{2}}]=2. \label{eq:ASD_condition}
\end{align}
The first equation is trivially satisfied.  The second equation is
\begin{align*}
 &[X^1,X^{\bar{1}}]+[X^2,X^{\bar{2}}]\nonumber\\
 &= U_l \left[ (N+2) f(N)^2 \theta(N+1\ge l) 
 - N f(N-1)^2 \theta(N-1\ge l) \right] U^\dagger_l=2.
\end{align*}
Because $U_l$ shifts $N$ by $l$, this condition is equivalent to
the recursive equation
\begin{align*}
 (N+2) f(N)^2  - N f(N-1)^2 = 2, \qquad N=l,l+1,\dots.
\end{align*}
Solving this equation, we obtain
\begin{align*}
 f_L(N)=\sqrt{\frac{(N-l+1)(N+l+2)}{(N+1)(N+2)}}\;\theta(N\ge l),
\end{align*}
which is a special case of (\ref{eq:the_solution:D=4}) with $L=l$.
This is the same as the instanton solution obtained by Nekrasov and
Schwarz \cite{Nekrasov-Schwarz} (see also \cite{Furuuchi}) by the
non-commutative ADHM construction, and describes ${\cal
N}_2(l)=\frac{l(l+1)}{2}$ instantons on top of each other at the origin.

So far, the gauge group has not been specified, but
$X_0^\alpha=f(N)a^\alpha$ has always been proportional to the unit
matrix.  However, in this $D=4$ ($d=2$) case, the ansatz can be
generalized to $U(2)$ gauge group:
\begin{align*}
 X^{\alpha}&=\left(
 \begin{matrix}
  U_l f_1(N) a^{\alpha} U^\dagger_l & U_l \epsilon^{\alpha\beta} a^{\bar{\beta}} g(N) \\
  0 & f_2(N) a^{\alpha}
 \end{matrix}
 \right),
\end{align*}
where $\epsilon^{12}=-\epsilon^{21}=1$.
For this ansatz, the anti-self-duality condition (\ref{eq:ASD_condition})
reads
\begin{align*}
 (N+2) f_1(N)^2 - N f_1(N-1)^2 + N g(N-1)^2 &=2,\quad N=l,l+1,\dots,\\
 (N+2) f_2(N)^2 - (N+2) g(N)^2 \theta(N\le l-1) \quad & \nonumber\\
 - N f_2(N-1)^2&=2, \quad N=0,1,2,\dots,\\
 (N+2) f_1(N) g(N) - N g(N-1) f_2(N-1)&=0,\quad N=l,l+1,\dots.
\end{align*}
{}From these recursive equations, it follows that
\begin{align*}
 f_1(0)=f_1(1)=\cdots=f_1(l-1)&=0,\\
 f_2(0)=f_2(1)=\cdots=f_2(l-2)&=1,\\
 g(0)=g(1)=\cdots=g(l-2)&=0,\\
 f_2(l-1)^2+g(l-1)^2&=1.
\end{align*}
Therefore we have one free parameter $f_2(l-1)$ (or equivalently,
$g(l-1)$).  This solution is the same as the instanton solution obtained
by ADHM construction and describes ${\cal N}_2(l)=\frac{l(l+1)}{2}$
instantons on top of each other at the origin, the free parameter
corresponding to the size of the instantons.

Lower level solutions are

\noindent
$l=1$:
\begin{align*}
 f_1(N)^2&=1-\frac{2+\rho^2}{(N+1)(N+2+\rho^2)},\\
 f_2(N)^2&=1+\frac{\rho^2}{(N+2)(N+3+\rho^2)},\\
 g(N)^2  &=\frac{\rho^2(2+\rho^2)}{(N+1)(N+2)(N+2+\rho^2)(N+3+\rho^2)},\\
 g(0)^2  &=\frac{\rho^2}{6+2\rho^2},
\end{align*}
\noindent
$l=2$:
\begin{align*}
 f_1(N)^2&=1-\frac{3(2+\rho^2)(N+3)}{(N+1)[(N+2)(N+3)+3\rho^2(N+1)]},\\
 f_2(N)^2&=1+\frac{3\rho^2 N}{(N+2)[(N+3)(N+4)+3\rho^2(N+2)]},\\
 g(N)^2  &=\tfrac{9\rho^2(2+\rho^2)N(N+3)}{(N+1)(N+2)[(N+2)(N+3)+3\rho^2(N+1)][(N+3)(N+4)+3\rho^2(N+2)]},\\
 g(1)^2  &=\frac{\rho^2}{20+9\rho^2}.
\end{align*}

The ansatz can be straightforwardly generalized to $U(k)$:
\begin{align*}
 X^{\alpha}&=\left(
 \begin{matrix}
  U_l f_1 a^{\alpha} U^\dagger_l & U_l \epsilon^{\alpha\beta} a^{\bar{\beta}} g_1 &&&\\
  & f_2 a^{\alpha} & \epsilon^{\alpha\beta} a^{\bar{\beta}} g_2 &&\\
  && \ddots & \ddots & \\
  &&& f_{k-1} a^{\alpha} & \epsilon^{\alpha\beta} a^{\bar{\beta}} g_{k-1} \\
  &&&& f_{k} a^{\alpha} 
 \end{matrix}
 \right),
\end{align*}
where $f$'s and $g$'s are functions of $N$.

\subsection{Mixing shift operator solitons and NS-type instantons}
\label{section:mixing_NS_and_HKL}

In this subsection, we go back to general $d$ and consider relaxing the
assumption $UU^\dagger=1$ in the NS-type instanton ansatz
(\ref{eq:ansatz}).  We will see that this corresponds to mixing shift
operator solitons and NS-type instantons.

If we only require $U^\dagger U=1-P$ and do not require $U U^\dagger =1$, then
$U U^\dagger$ is generally a projection operator
\begin{align*}
 U U^\dagger=1-P',\quad P'{}^\dagger=P',\quad P'{}^2=P'.
\end{align*}
We require
\begin{align*}
 \bra{\{n\}} U U^\dagger \ket{\{n\}} &\to 1
 \quad (N\to \infty)
\end{align*}
so $X^{\alpha}$ and $X^{\bar{\alpha}}$ approach respectively
$a^{\alpha}$ and $a^{\bar{\alpha}}$  at large $N$.  Hence we can
set
\begin{align*}
 P'=\theta(\{n\}\in{\cal P}'),
\end{align*}
where ${\cal P}'$ is a finite subset of ${\bf
N}^d=\{(n^1,\dots,n^d);n^1,\dots,n^d=0,1,2,\dots\}$.  Now $U$ is
essentially a shift operator which maps one to one the subspace
$(1-P){\cal H}$ onto the subspace $(1-P'){\cal H}$.  We write this $U$
operator as $U_{l}^{(m)}$ henceforth, where $m$ is the number of
elements of ${\cal P}'$.

With this change, the ansatz (\ref{eq:ansatz}) can be generalized to
\footnote{Just as stated in footnote
\ref{footnote:position_moduli_of_shift_operator_solitons}, there exists
the additional freedom of adding finite matrices in the subspace
$P'{\cal H} $ to (\ref{eq:generalized_ansatz}). However we do not
consider this generalization for simplicity.}
\begin{align}
 X^\alpha 
 &=  U_{l}^{(m)} f(N) a^\alpha U_l^{(m)}{}^\dagger 
 = U_l^{(m)} X_0^\alpha U_l^{(m)}{}^\dagger, \nonumber\\
 X^{\bar{\alpha}} 
 &= U_{l}^{(m)} a^{\bar{\alpha}} f(N) U_{l}^{(m)}{}^\dagger
 = U_l^{(m)} X_0^{\bar{\alpha}} U_l^{(m)}{}^\dagger.
 \label{eq:generalized_ansatz}
\end{align}
In order for this $X$ to satisfy the equation of motion, $f(N)$ should still
be given by (\ref{eq:the_solution}).  The field strength is modified
from (\ref{eq:field_strength_of_ansatz-1}) to
\begin{align}
 F_{\alpha\bar{\beta}}
 &= U_{l}^{(m)} (F_0)_{\alpha\bar{\beta}} U_{l}^{(m)}{}^\dagger
 \MP i \delta^{\bar{\alpha}\beta} P',
 \label{eq:field_strength_with_P'}
\end{align}
where $(F_0)_{\alpha\bar{\beta}}$ is still given by
(\ref{eq:field_strength_of_ansatz-2}).  The values of the action
(\ref{eq:value_of_action}) and the topological charge
(\ref{eq:value_of_topological_charge}) are modified as
\begin{align}
 S
 &=(2\pi)^d \frac{d}{2}\
 [d\, ({\cal N}_d(L)+ m)-{\cal N}_d(l)],\nonumber\\
 Q
 &= \begin{cases}
     {\cal N}_d(l) - m & (d\ge 2) , \\
     -(L-l) - m & (d=1).
    \end{cases} \label{eq:modified_value_of_action_and_tpl_charge}
\end{align}
Therefore, introducing $P'$ changes the action and
the topological charge by $m$.

As mentioned earlier, this generalization (\ref{eq:generalized_ansatz})
amounts to adding $m$ shift operator solitons (\ref{eq:shift_operator_solitons})
to the NS-type instanton. This is clear because the new solution
(\ref{eq:generalized_ansatz}) can be obtained from the old solution
(\ref{eq:ansatz}) by an ``almost gauge'' transformation:
\begin{align}
 U_{l}^{(m)} X_0 U_{l}^{(m)}{}^\dagger
 &= V^\dagger (U_l X_0 U_l^\dagger) V,\\
 VV^\dagger &=1, \quad V^\dagger V=1-P'.
\end{align}


Note that the shift operator soliton gives an opposite sign contribution to the
topological charge (\ref{eq:modified_value_of_action_and_tpl_charge}).
This is consistent with the fact that in the $D=4$ case a shift operator
soliton can be obtained by taking the radius $\rho\to 0$ limit of a
$U(2)$ anti-self-dual instanton obtained via the ADHM construction
\cite{Furuuchi} in the case of anti-self-dual non-commutativity.
On the other hand, because we are taking self-dual
non-commutativity, our solution, which becomes anti-self-dual when $L=l$,
should have  opposite topological charge as compared to shift operator solitons.

\bigskip 
In the $D=2$ ($d=1$) case, we can take for example
\begin{align}
 U=\sum_{n^1=0}^\infty \ket{n^1+m} \bra{n^1+l}, \quad m,l\ge 0,
 \label{eq:generalized_shift_operator_in_2-D}
\end{align}
which satisfies
\begin{align}
 U^\dagger U=1-P_{l},
 \quad
 U U^\dagger=1-P_m.  \label{eq:generalized_UU_conditions_in_2D}
\end{align}
Using (\ref{eq:the_solution:D=2}), one can show
\begin{align*}
 X^1=U f_L(N) a^1 U^\dagger = U^\dagger_{L-l+m} a^1 U_{L-l+m} .
\end{align*}
This is again consistent with the result \cite{Gross-Nekrasov} concerning  the
most general soliton solution in 2-dimensional pure NCYM.

In the $D=4$ ($d=2$) case, it is clear that the
field strength (\ref{eq:field_strength_with_P'}) can never be
anti-self-dual for non-vanishing $P'$.  This can
be also seen  in terms of relations following from 
(\ref{eq:modified_value_of_action_and_tpl_charge}):
\begin{align*}
 \frac{S}{(2\pi)^2}+Q&=2{\cal N}_d(L)+m,\nonumber\\
 \frac{S}{(2\pi)^2}-Q&=2[{\cal N}_d(L)-{\cal N}_d(l)]+3m.
\end{align*}
The right hand sides never become zero unless $m=0$.

\section{Seiberg--Witten map of NCYM solitons}

In this section, we consider the SW map (\ref{eq:operator_SW_map}) of
solitonic solutions in NCYM.  We derive a differential equation
describing the evolution of fields under the SW map.  The shift operator
soliton is shown to be invariant under this flow.
For the NS-type instanton we first expand the solution in a power series.
Based partly on numerical analysis, we solve for the coefficients,
resum the series, and so obtain the commutative description.  
Both shift operator solitons and NS-type instantons are singular at the 
origin  in
commutative variables.  Furthermore, both have zero size as the 
original noncommutativity parameter is taken to zero.

\subsection{Shift operator soliton}

Let us consider the SW map of the shift operator soliton
(\ref{eq:shift_operator_solitons}):
\begin{align}
 \OP{X}^i(\theta)=\OP{U}^\dagger \OP{x}^i(\theta) \OP{U},
 \qquad \OP{U}\OP{U}^\dagger=1, \quad \OP{U}^\dagger \OP{U}=1-\OP{P}.
 \label{eq:shift_operator_solitons_with_theta_explicit}
\end{align}
By inserting this into (\ref{eq:operator_SW_map}), we find
\begin{align*}
 \OP{X}^i(\theta+\delta\theta)
 = \OP{U}^\dagger 
 \left(\OP{x}^i(\theta)+\frac{1}{2}\delta\theta^{ij}\theta_{jk}\OP{x}^{k}(\theta)\right) \OP{U}
 = \OP{U}^\dagger \OP{x}^i(\theta+\delta\theta) \OP{U}.
\end{align*}
Therefore, the shift operator soliton is invariant under the SW map.  

For non-commutativity of the form
(\ref{eq:homogeneous_non-commutativity}) and for
$\OP{P}=|\vec{0}\rangle\langle\vec{0}|$, the inverse Weyl transformation of the
field strength (\ref{eq:field_strength_of_shift_operator_solitons}) is
\begin{align*}
 F_{\alpha\bar{\beta}}(x)
 &=
 -\frac{i\delta^{\bar{\alpha}\beta}}{\theta} \!
 \int\!\frac{{\rm d}^{2d} k}{(2\pi)^{2d}}\, (2\pi\theta)^d \Tr[\OP{P}e^{-ik\cdot \OP{x}(\theta)}] e^{ik\cdot x}
 = 
 -\frac{2^d i \delta^{\bar{\alpha}\beta}}{\theta} e^{-r^2/\theta},
\end{align*}
where $r^2 \equiv \sum_{i=1}^D (x^i)^2 = 2 \bar{z}^{\bar{\alpha}} z^\alpha$.
Therefore, in the $\theta\to 0$ limit, we obtain the commutative
description of the shift operator solitons:
\begin{align*}
 F=F^2=\cdots=F^{d-1}=0,\quad
 F^d \propto \delta^{(2d)}(x),
\end{align*}
which is consistent with the result obtained by direct calculation using
the exact form of the SW map \cite{K.Hashimoto-Ooguri}.

\subsection{NS-type instanton}

With non-commutativity parameter 
(\ref{eq:homogeneous_non-commutativity}), if the solution is
of the form
\begin{align}
 \OP{X}^\alpha(\theta)
 &=
 \OP{U} f(\OP{N};\theta) \, \OP{x}^\alpha(\theta) \OP{U}^\dagger, \\
 \label{eq:ns-type-solution-with-any-theta}
 \OP{U}^\dagger \OP{U}=1-\OP{P},\quad
 \OP{U} \OP{U}^\dagger&=1-\OP{P}',\quad
 f(\OP{N};\theta) \OP{P} = \OP{P} f(\OP{N};\theta) =0,
\end{align}
then the SW map (\ref{eq:operator_SW_map}) is
\begin{align}
 \delta\OP{X}^\alpha(\theta)
 &=
 \frac{\delta\theta}{2\theta} \OP{U} f(\OP{N};\theta)
 [ (\OP{N}+1) f(\OP{N};\theta)^2 - \OP{N} f(\OP{N}-1;\theta)^2 ] 
 \OP{x}^\alpha(\theta)\OP{U}^\dagger.
 \label{eq:SW_map_of_NS-type_instanton-RHS}
\end{align}
This is again of the form of (\ref{eq:ns-type-solution-with-any-theta}),
and thus we obtain an equation which describes the evolution of the
solution under the SW map:
\begin{align*}
 \frac{\partial f(\OP{N};\theta)}{\partial \theta}
 =
 \frac{f(\OP{N};\theta)}{2\theta}
 [ (\OP{N}+1) f(\OP{N};\theta)^2 - \OP{N} f(\OP{N}-1;\theta)^2 -1] ,
\end{align*}
where the last term comes from rewriting $\OP{x}(\theta+\delta\theta)$
in terms of $\OP{x}(\theta)$.  Defining $t\equiv \ln (\theta/\theta_0)$
this equation can be rewitten as
\begin{align}
 \frac{\partial f(\OP{N};t)^2} {\partial t}
 = f(\OP{N};t)^2
 [ (\OP{N}+1) f(\OP{N};\theta)^2 - \OP{N} f(\OP{N}-1;\theta)^2 -1].
 \label{eq:SW_map_of_NS-type_solution-f(N;t)^2}
\end{align}

The above nonlinear differential equation is difficult to solve directly. 
We will attack it by expanding $f$ in a power series and solving for the
coefficients.  
It is simplest to expand the the solution as
\begin{align}
 f(\OP{N};t)^2=1+\frac{c_1(t)}{\OP{N}+1}+\frac{c_2(t)}{(\OP{N}+1)^2}+\dots
=\sum_{n=0}^\infty \frac{c_n(t)}{(\OP{N}+1)^n}.
\quad c_0(t)\equiv 1,
 \label{eq:1/(N+1)-expansion_of_f^2}
\end{align}
Plugging this into (\ref{eq:SW_map_of_NS-type_solution-f(N;t)^2}) we obtain
\begin{align}
 \frac{{\rm d}c_n(t)}{{\rm d}t}
 =
 - \sum_{p=0}^{n-2} \sum_{q=0}^{p}
 \left( {n-q-1 \atop  p-q} \right)
 c_{q}(t) c_{p-q+2}(t) 
 \qquad (n\ge 1),
 \label{eq:D.E._for_c_n(t)}
\end{align}
where $({p \atop q})\equiv\frac{p!}{q!(p-q)!}$.
By induction, it is not hard to show that the solution
for the coefficients  $c_n(t)$ is of the form
\begin{align}
 c_n(t)=\sum_{m=1}^{n-1} c_n^{(m)}e^{-mt}
 \qquad (n\ge 2),
 \label{eq:c_n=sum_c_n^(m)}
\end{align}
except for $c_1(t)=\text{const}$.  The coefficients $c_n^{(m)}$, $n\ge
2$, $1\le m \le n-2$ are determined recursively by
\begin{multline}
 c_n^{(m)} = -\frac{1}{n-m-1} \Bigg[
 \sum_{p=m+1}^{n-1} \left( { n-1 \atop p-2 } \right) c_p^{(m)}\\
 + \sum_{t,q\ge 0}^{t+q\le n-4} \sum_{r=1}^{q+1} \sum_{s=1}^{t+1}
 \left( { n-q-3 \atop t } \right) c_{q+2}^{(r)} c_{t+2}^{(s)} \delta_{r+s,m}
\Bigg]
 \label{eq:recursive_formula_for_c_n^(m)}
\end{multline}
and $c_n^{(n-1)}$ is determined by the initial condition
$c_n(t=0) = \sum_{m=1}^{n-1} c_n^{(m)}$.

Now let us consider evaluating the $\theta\to 0$ limit of the c-number
function $F_{\alpha\bar{\beta}}(x;\theta)$ for $x\neq 0$.  In terms of
$f(\OP{N};t)$, the operator $\OP{F}_{\alpha\bar{\beta}}(\theta)$ can be
written as
\begin{align*}
 \OP{F}_{\alpha\bar{\beta}}(\theta)
 =
 \frac{i}{\theta} \OP{U} 
 \left[
 f(\OP{N};t)^2 \delta^{\bar{\alpha}\beta}
 + (f(\OP{N};t)^2-f(\OP{N}-1;t)^2)\OP{a}^{\bar{\alpha}}\OP{a}^{\beta}
 \right]
 \OP{U}^\dagger
 -\frac{i}{\theta}\delta^{\bar{\alpha}\beta}.
\end{align*}
Roughly, $\OP{\bar{z}}^{\bar{\alpha}} \OP{z}^{\alpha} = \theta \OP{N}$
in operator language corresponds to $\bar{z}^{\bar{\alpha}}
z^{\alpha}\equiv\bar{z}z$ in c-number function language, therefore
$\theta\to 0$ implies $\OP{N} \simeq \bar{z}z/\theta \to \infty$ as long
as $\bar{z}z\neq 0$.  In this $\OP{N}\to \infty$ limit,
$F_{\alpha\bar{\beta}}(x;\theta)$ can be obtained from
$\OP{F}_{\alpha\bar{\beta}}(\theta)$ simply by substituting $\OP{N}$ and
$\OP{a}^{\bar{\alpha}}$ with $\bar{z}z/\theta$ and
$z^{\alpha}/\sqrt{\theta}$, respectively:
\begin{align*}
 F_{\alpha\bar{\beta}}(x;\theta)
 & \stackrel{\substack{\theta\sim 0\\ \bar{z}z\neq 0 \vspace{.5ex}}}{\sim}  
 \frac{i}{\theta} 
 \left[
 f(\tfrac{\bar{z}z}{\theta};t)^2 \delta^{\bar{\alpha}\beta}
 + (f(\tfrac{\bar{z}z}{\theta};t)^2-f(\tfrac{\bar{z}z}{\theta}-1;t)^2)
 \frac{\bar{z}^{\bar{\alpha}}z^{\beta}}{\theta}
 \right]
 -\frac{i}{\theta}\delta^{\bar{\alpha}\beta}\\
 & ~~\sim~
 \frac{i}{\theta} 
 \left[
 (f(\tfrac{\bar{z}z}{\theta};t)^2-1) \delta^{\bar{\alpha}\beta}
 + \frac{\partial f(\tfrac{\bar{z}z}{\theta};t)^2}{\partial (\tfrac{\bar{z}z}{\theta})}
 \frac{\bar{z}^{\bar{\alpha}}z^{\beta}}{\theta}
 \right].
\end{align*}
In the first line, $\OP{U}$ and $\OP{U}^\dagger$ operators are not
needed because finite shifts in the Hilbert space introduced by these
operators cannot change $F_{\alpha\bar{\beta}}(x;\theta)$ at finite
$\bar{z}z$ in the $\OP{N}\simeq \bar{z}z/\theta \to\infty$ limit.  They
might change $\OP{N}$ into $\OP{N}+1$, $\OP{N}-2$, etc.\ on some state,
but this is irrelevant in the $\OP{N}\to\infty$ limit.

Using the expansions
(\ref{eq:1/(N+1)-expansion_of_f^2}) and (\ref{eq:c_n=sum_c_n^(m)}),
\begin{align*}
 F_{\alpha\bar{\beta}}(x;\theta)
 & \stackrel{\substack{\theta\sim 0\\ \bar{z}z\neq 0 \vspace{.5ex}}}{\sim}  
 \frac{i}{\theta}
 \sum_{n=1}^{\infty} \sum_{m=1}^{n-1}
 \frac{\theta^{n-m}\theta_0^m}{(\bar{z}z)^{n}}
 \left[
 \delta^{\bar{\alpha}\beta}-n\frac{\bar{z}^{\bar{\alpha}}z^\beta}{\bar{z}z}
 \right]
 c_n^{(m)}\\
 & \to
 \frac{i}{\theta_0}
 \sum_{n=1}^{\infty}
 \left( \frac{\theta_0}{\bar{z}z} \right)^n
 \left[
 \delta^{\bar{\alpha}\beta}-n\frac{\bar{z}^{\bar{\alpha}}z^\beta}{\bar{z}z}
 \right]
 c_n^{(n-1)}  \qquad  (\theta\to 0),
\end{align*}
assuming that we can interchange the order of summation and limit.
Therefore, the commutative description away from the origin is given by
\begin{align}
 F_{\alpha\bar{\beta}}(x;\theta=0)
 = -\frac{2i}{\theta_0}
 \left[
 h(\rho) \delta^{\bar{\alpha}\beta}+ 
 \frac{h'(\rho)}{\rho}
 \frac{\bar{z}^{\bar{\alpha}}z^\beta}{\theta_0}
 \right]
 \qquad (x\neq 0),
 \label{eq:commutative_field_strength_in_terms_of_c_n^(n-1)}
\end{align}
where $\rho^2 \equiv r^2/\theta_0 = 2\bar{z}z/\theta_0$ and\footnote{We
introduce the function $h$ for comparison with the discussion in
section 4.2 of Ref.\ \cite{Seiberg-Witten}.}
\begin{align}
 h(\rho)\equiv -\frac{1}{2} \sum_{n=1}^\infty c_n^{(n-1)} 
 \left( \frac{2}{\rho^2} \right)^n.
 \label{eq:h_func}
\end{align}
Note that only the coefficients $c_n^{(n-1)}$, which are determined only
by the initial condition $f(\OP{N},\theta_0)$, contribute in the $\theta
\rightarrow 0$ limit.  In particular, the $\OP{U}$ and $\OP{U}^\dagger$
operators cannot change the commutative description away from the
origin, although they might contribute to the singularity at the origin.

The topological charge density, calculated from Eq.\
(\ref{eq:commutative_field_strength_in_terms_of_c_n^(n-1)}) using
the commutative product, is
\begin{align*}
 \sigma(x)=-\frac{1}{(\pi\theta_0)^d}
 [h^d+(\rho/2) h' h^{d-1}], \qquad
 Q=\int {\rm d}^D x\, \sigma(x).
 \label{eq:commutative_top_charge_density_in_terms_of_Phi}
\end{align*}
Integrating $\sigma(x)$ over ${\bf R}^D-\{0\}$, we obtain
\begin{align}
 Q'\equiv\int_{x \neq 0} {\rm d}^D x\, \sigma(x)
 =-\frac{1}{d!}\Big[\rho^{2d} h^d \Big]_{\rho=0}^{\rho=\infty}.
\end{align}

\bigskip Specifically, let us consider the case with $d=2$ ($D=4$),
$\OP{U}=\OP{U}_l^{(m)}$, $L=l=1$, 
which includes the Nekrasov--Schwarz instanton ($m=0$).
The coefficients $c_n^{(m)}$ are determined recursively by Eq.\
(\ref{eq:recursive_formula_for_c_n^(m)}) along with the initial
condition $c_1(t=0)=0$ and $c_n(t=0)=2(-1)^{n+1}$, $n\ge 2$.  
We have been unable to solve these recursion relations analytically.
Nevertheless, explicit calculations up  to $n=100$ yield quite a 
simple result.  Based on this, we believe that the solution is
\begin{align}
 c_n^{(n-1)}(t)=
 \begin{cases}
  -(-4)^m  & n=4m   ,\quad m=1,2,3,\dots, \\
  -2(-4)^m & n=4m+1 ,\quad m=1,2,3,\dots, \\
  -2(-4)^m & n=4m+2 ,\quad m=0,1,2,\dots,\\
  0        & \text{otherwise},
 \end{cases}
 \label{eq:c_n^(n-1)_for_d=2,L=1}
\end{align}
which gives
\begin{align}
 h(\rho)=\frac{4(\rho^2-4)}{\rho^2(\rho^4-4\rho^2+8)}.
\end{align}
Therefore the field strength away from the origin is
\begin{multline}
 F_{\alpha\bar{\beta}}(x;\theta=0)
 = -\frac{i}{\theta_0}
 \Bigg[
 \frac{8(\rho^2-4)}{\rho^2(\rho^4-4\rho^2+8)}\delta^{\bar{\alpha}\beta}\\
 -
 \frac{32(\rho^6-8\rho^4+16\rho^2-16)}{\rho^4(\rho^4-4\rho^2+8)^2}\frac{\bar{z}^{\bar{\alpha}}z^\beta}{\theta_0}
 \Bigg] 
 \qquad (x\neq 0).
\label{FS}
\end{multline}

The field strength $F$ has significant nonzero values in a region
$r \lesssim \sqrt{\theta_0}$ and the region shrinks
and vanishes in the $\theta_0 \to 0$ limit, leaving a singularity at the
origin.  This is consistent with the fact that there are no smooth
instanton solutions in commutative $U(1)$ YM theory, which $U(1)$ NCYM
theory approaches in the $\theta_0 \to 0$ limit.  The asymptotic
behavior of the field strength is $F\sim 1/r^4$ for $r\to\infty$.  This
behavior is the same as that of the solitonic solution to the
nonpolynomial action obtained by applying the Seiberg--Witten zero slope
limit to the Born--Infeld action \cite{Seiberg-Witten,Terashima}.  This
nonpolynomial action agrees with the SW map of the NCYM action up to
derivative corrections.  The agreement in the asymptotic forms of the
respective solutions is consistent with the fact that the ignored higher
order terms becomes irrelevant at large distance.  The $r\to 0$
behaviors also match and are $F\sim 1/r^2$; however we cannot
rationalize this as above since the higher order terms are no longer
negligible at short distance.  Instead, the behavior $F\sim 1/r^2$ is
just what is needed to give the solution a finite and nonzero
topological charge.

The topological charge density is, from Eq.\
(\ref{eq:commutative_top_charge_density_in_terms_of_Phi}),
\begin{align*}
 \sigma(x)=
 \frac{1}{(\pi\theta_0)^2}
 \frac{16(\rho^6-12\rho^4+40\rho^4-32)}{\rho^2(\rho^8-4\rho^4+8)^3}
 \qquad(x\neq 0).
\end{align*}
Integrating $\sigma(x)$ over ${\bf R}^4-\{0\}$ gives
\begin{align}
 Q'=
 \int_{x\neq 0}{\rm d}^4 x \, \sigma(x)
 = -\frac{1}{2}[\rho^4 h^2]^{\rho=\infty}_{\rho=0}
 =2.
\end{align}
As we have stressed, our derivation of the commutative form of the instanton
is only valid away from the origin.  The behavior at the origin can now
be determined. 
Since we know that the full topological charge $Q=\int {\rm d}^4x\,
\sigma(x)$ should be equal to ${\cal N}_2(1)-m=1-m$ from Eq.\
(\ref{eq:modified_value_of_action_and_tpl_charge}), we should be able to
extend $\sigma(x)$ to include the singularity at the origin:
\begin{align*}
 \sigma(x)=
 \frac{1}{(\pi\theta_0)^2}
 \frac{16(\rho^6-12\rho^4+40\rho^4-32)}{\rho^2(\rho^8-4\rho^4+8)^3}
 -(m+1)\delta^{(4)}(x)
 \quad \text{(for all $x$)}.
\end{align*}
Since $m\ge 0$, we always have to add a delta function singularity at
the origin.  In particular, $m=0$ corresponds to the original
Nekrasov--Schwarz instanton.

To summarize, we have found the commutative description of the 
Nekrasov--Schwarz instanton by solving the infinitesimal 
SW map in operator form.  Based  partly on numerical analysis,
the commutatative field strength was found to be (\ref{FS}), except
at the origin where an extra delta function contribution is needed
to obtain the correct topological charge.  The commutative 
description has the expected property of shrinking to zero size as
the original noncommutativity parameter $\theta_0$ is taken to zero.
It would of course be desirable to prove our formula 
(\ref{eq:c_n^(n-1)_for_d=2,L=1}),  and to find similar commutative
descriptions of the other instanton solutions found in this paper.

\section*{Acknowledgment}
 Work supported by NSF grant PHY-0099590.   

\appendix

\section{Actions and topological charges of NS-type instantons}
\label{section:action_and_Chern_chars}

In this appendix, we compute the action and the topological charge
($d$-th Chern character) of NS-type instantons.

The action can be evaluated as
\begin{align}
 \frac{S}{(2\pi)^d}
 &=-\frac{1}{2}\Tr\, F_{\alpha\bar{\beta}} F_{\beta\bar{\alpha}}
 =-\frac{1}{2}\Tr\, 
 (F_0)_{\alpha\bar{\beta}} (F_0)_{\beta\bar{\alpha}} 
 \nonumber\\
 &=\frac{1}{2}\Tr
 \Big[
 \delta^{\bar{\alpha}\beta} \delta^{\bar{\beta}\alpha}\theta(l\le N\le L)\nonumber\\
 &\qquad\qquad+
 F(N)^2
 \theta(N\ge L)(N\delta^{\bar{\alpha}\beta}-d\, a^{\bar{\alpha}} a^{\beta})
 (N\delta^{\bar{\beta}\alpha}-d\, a^{\bar{\beta}} a^{\alpha})
 \Big]
 \nonumber\\
 &=\frac{d}{2}\sum_{N=l}^{L-1}
 {\cal D}_d(N)+\frac{1}{2}\sum_{N=L}^{\infty}
 F(N)^2
 d(d-1)N(N+d){\cal D}_d(N)\nonumber\\
 &=
 \frac{d}{2} [{\cal N}_d(L)-{\cal N}_d(l)]
 + \frac{d(d-1)}{2} {\cal N}_d(L)\nonumber\\
 &= \frac{d}{2} [d\,{\cal N}_d(L)-{\cal N}_d(l)].
 \label{eq:appendix:calc_of_action}
\end{align}
Here 
\begin{align*}
 F(N)&\equiv\frac{L(L+1)\dots(L+d-1)}{N(N+1)\dots(N+d)},\\
 {\cal D}_d(N)&\equiv\frac{(N+1)(N+2)\dots(N+d-1)}{(d-1)!},\\
 {\cal N}_d(L)&\equiv\sum_{N=0}^{L-1} {\cal D}_d(N) = \frac{L(L+1)\dots(L+d-1)}{d!}.
\end{align*}
${\cal D}_d(N)$ is the number of states with $N=n^1+\dots+n^d$, and
${\cal N}_d(L)$ is the number of states with $N\le L-1$.  The second
equality in (\ref{eq:appendix:calc_of_action}) is understood as the
following.  As can be seen from the explicit form
(\ref{eq:field_strength_of_ansatz-1}), $F_0$ kills $P_l$ on its left and
right. Hence a $U^\dagger_l U_l=1-P_l$ operator between any two $F_0$'s
can be replaced with a unit operator.

The calculation of the Chern character is more involved.
Since the only non-vanishing component of the field strength is
$F_{\alpha\bar{\beta}}$, we find, 
\begin{align}
 Q &\equiv
 \frac{-1}{(\MPf 2\pi)^d d!} \int \wedge^d F\nonumber\\
 &= 
 \frac{-1}{(\MPf 2\pi)^d d!}  \int 
 (F_{\alpha_1\bar{\beta}_1} {\rm d}z^{\alpha_1} \wedge {\rm d}\bar{z}^{\bar{\beta}_1} )
 \wedge \cdots \wedge
 (F_{\alpha_d\bar{\beta}_d} {\rm d}z^{\alpha_d} \wedge {\rm d}\bar{z}^{\bar{\beta}_d} )\nonumber\\
 &= 
 \frac{-1}{(\MPf 2\pi)^d d!} \int 
 \epsilon^{\alpha_1 \dots \alpha_d}
 \epsilon^{\bar{\beta}_1 \dots \bar{\beta}_d}
 F_{\alpha_1\bar{\beta}_1} \cdots F_{\alpha_d\bar{\beta}_d}
 {\rm d}z^{1} \wedge {\rm d}\bar{z}^{\bar{1}}
 \wedge \cdots \wedge
 {\rm d}z^{d} \wedge {\rm d}\bar{z}^{\bar{d}}\nonumber\\
 &= 
 \frac{-1}{(\MPf 2\pi)^d d!} \int 
 \epsilon^{\alpha_1 \dots \alpha_d}
 \epsilon^{\bar{\beta}_1 \dots \bar{\beta}_d}
 F_{\alpha_1\bar{\beta}_1} \cdots F_{\alpha_d\bar{\beta}_d}
 \frac{{\rm d}^D x}{i^d}
 \nonumber\\
 &= 
 \tfrac{1}{-(\MPf i)^d d!} 
 \epsilon^{\alpha_1 \dots \alpha_d}
 \epsilon^{\bar{\beta}_1 \dots \bar{\beta}_d} 
 \Tr\, 
 F_{\alpha_1\bar{\beta}_1} \cdots F_{\alpha_d\bar{\beta}_d}
 \nonumber\\
 &= 
 \tfrac{1}{-(\MPf i)^d d!} 
 \epsilon^{\alpha_1 \dots \alpha_d}
 \epsilon^{\bar{\beta}_1 \dots \bar{\beta}_d} 
 \Tr\, 
 (F_0)_{\alpha_1\bar{\beta}_1} \cdots (F_0)_{\alpha_d\bar{\beta}_d},
 \label{eq:appendix:Charn_char}
\end{align}
where summation over identical upper and lower indices is implied. The
totally antisymmetric $\epsilon$ symbol is defined as $\epsilon^{12\dots
d}=\epsilon^{\bar{1}\bar{2}\dots \bar{d}}=1$.  Plugging in the explicit
form of $(F_0)_{{\alpha}\bar{\beta}}$,
\begin{align}
 &\hspace{-3ex}
 \epsilon^{\alpha_1 \dots \alpha_d}
 \epsilon^{\bar{\beta}_1 \dots \bar{\beta}_d} 
 (F_0)_{\alpha_1\bar{\beta}_1} \cdots (F_0)_{\alpha_d\bar{\beta}_d} \nonumber\\
 =&
 (\MPf i)^d \big[
 \epsilon^{\alpha_1 \dots \alpha_d}
 \epsilon^{\bar{\beta}_1 \dots \bar{\beta}_d} 
 \delta^{\bar{\alpha}_1 \beta_1} \cdots \delta^{\bar{\alpha}_d \beta_d} \theta(l\le N \le L-1)
 + F(N)^d \theta(N\ge L) \nonumber\\
 & \qquad\qquad\quad\times
 \epsilon^{\alpha_1 \dots \alpha_d}
 \epsilon^{\bar{\beta}_1 \dots \bar{\beta}_d}
 (N\delta^{\bar{\alpha}_1 \beta_1} -d\, a^{\bar{\alpha}_1} a^{\beta_1} )
 \dots
 (N\delta^{\bar{\alpha}_d \beta_d} -d\, a^{\bar{\alpha}_d} a^{\beta_d} )
 \big] 
 \nonumber\\
 =&
 (\MPf i)^d \Big[
 d! \, \theta(l\le N\le L-1)
 \nonumber \\
 &\qquad\quad
 + F(N)^d \theta(N\ge L) \sum_{k=0}^d \left({d \atop k}\right)
 \epsilon^{\alpha_1 \dots \alpha_{k} \alpha_{k+1} \dots \alpha_d}
 \epsilon^{\bar{\beta}_1 \dots \bar{\beta}_{k} \bar{\beta}_{k+1} \dots \bar{\beta}_d} \nonumber \\
 &\qquad\qquad\quad
 \times (-d)^k (a^{\bar{\alpha}_1}a^{\beta_1}) \cdots (a^{\bar{\alpha}_k}a^{\beta_k})
 (N\delta^{\bar{\alpha}_{k+1}\beta_{k+1}}) \cdots (N\delta^{\bar{\alpha}_{d}\beta_{d}})
 \Big] \nonumber\\
 =&
 (\MPf i)^d \Big[
 d! \, \theta(l\le N\le L-1)
 + F(N)^d \theta(N\ge L) \nonumber \\
 & \qquad\quad
 \times \sum_{k=0}^d \left({d \atop k}\right)
 (d-k)!\, (-d)^k N^{d-k}
 \delta^{\alpha_1 \dots \alpha_k;\, \bar{\beta}_1 \dots \bar{\beta}_k}
 (a^{\bar{\alpha}_1}a^{\beta_1}) \cdots (a^{\bar{\alpha}_k}a^{\beta_k})
 \Big] \label{eq:epsilon-epsilon-F_0-F_0},
\end{align}
where 
\begin{align*}
 F(N)\equiv \frac{L(L+1)\cdots(L+d-1)}{N(N+1)\cdots(N+d)}
\end{align*}
and we have used the relation
\begin{align*}
 \epsilon^{\alpha_1 \dots \alpha_{k} \alpha_{k+1} \dots \alpha_d}
 \epsilon^{\bar{\beta}_1 \dots \bar{\beta}_{k} \bar{\alpha}_{k+1} \dots \bar{\alpha}_d} 
 &= (d-k)!\, 
 \delta^{\alpha_1 \dots \alpha_k; \, \bar{\beta}_1 \dots \bar{\beta}_k}
 ,\\
 \delta^{\alpha_1 \dots \alpha_k; \, \bar{\beta}_1 \dots \bar{\beta}_k}
 &\equiv 
 k! \times
 \left(
 \begin{array}{cc}
 \text{antisymmetrization of 
 $\delta^{\alpha_1 \bar{\beta}_1} \cdots \delta^{\alpha_k \bar{\beta}_k}$}\\
 \text{with respect to $\bar{\beta}_1 \dots \bar{\beta}_k$}
 \end{array}
 \right)
 .
\end{align*}
Furthermore, using the relation
\begin{align*}
 &\hspace{-5ex}
 \delta^{\alpha_1 \dots \alpha_k ;\, \bar{\beta}_1 \dots \bar{\beta}_k}
 (a^{\bar{\alpha}_1}a^{\beta_1}) \cdots (a^{\bar{\alpha}_k}a^{\beta_k}) \nonumber\\
 &= (-1)^{k-1}(d-1)(d-2)\cdots(d-k+1)N,\qquad k\ge 1,
\end{align*}
(\ref{eq:epsilon-epsilon-F_0-F_0}) becomes
\begin{align*}
 &\hspace{-3ex}
 \epsilon^{\alpha_1 \dots \alpha_d}
 \epsilon^{\bar{\beta}_1 \dots \bar{\beta}_d} 
 (F_0)_{\alpha_1\bar{\beta}_1} \cdots (F_0)_{\alpha_d\bar{\beta}_d} \nonumber\\
 =&
 (\MPf i)^d \Big[
 d! \, \theta(l\le N\le L-1) \nonumber\\
 &\qquad\qquad
 + F(N)^d \theta(N\ge L) 
 (d-1)!\,(d\,N^d+N^{d+1}-N(N+d)^d)
 \Big]. 
\end{align*}
Putting this back into (\ref{eq:appendix:Charn_char}), we obtain the
final result
\begin{align*}
 Q
 &= \frac{-1}{d!}
 \Big[
 d!\sum_{N=l}^{L-1} {\cal D}_d(N)
 \nonumber\\
 &\qquad\qquad\quad +(d-1)!
  \sum_{N=L}^{\infty} F(N)^d (d\,N^d+N^{d+1}-N(N+d)^d) {\cal D}_d(N)
 \Big]\nonumber\\
 &= \begin{cases}
     \frac{-1}{d!}
     \Big[
     d!\, ({\cal N}_d(L)-{\cal N}_d(l)) - (d-1)! \cdot d\, {\cal N}_d(L) \Big] & (d\ge 2), \\[1ex]
     \frac{-1}{d!}
     \Big[ d!\, ({\cal N}_d(L)-{\cal N}_d(l)) + 0 \Big] & (d=1)
    \end{cases}\nonumber\\
 &= \begin{cases}
     {\cal N}_d(l) & (d\ge 2), \\
     -(L-l) & (d=1).
    \end{cases}
\end{align*}

\section{Two Seiberg--Witten maps are related by unitary
     transformation}
     \label{sec:SW_and_SW'_different_by_unitary_transf}

In this appendix, we will show that solutions to the two SW maps
(\ref{eq:SW_map_of_X_operator_with_`gauge'_left}) and
(\ref{eq:operator_SW_map}), namely,
\begin{align}
 \frac{{\rm d} X_g^i}{{\rm d}\tau}
 =
 \frac{i}{4} \frac{{\rm d}\theta^{kl}}{{\rm d}\tau}
 \theta_{km}\theta_{ln} 
 \{X_g^m,[X_g^n,X_g^i]\}
 +i[g(X_g,x),X_g^i]
 \label{eq:SW_map_for_X_g}
\end{align}
and
\begin{align}
 \frac{{\rm d} X_0^i}{{\rm d}\tau}&=  
 \frac{i}{4} \frac{{\rm d}\theta^{kl}}{{\rm d}\tau}
 \theta_{km}\theta_{ln} 
 \{X_0^m,[X_0^n,X_0^i]\}
 \phantom{+i[g(X_g,x),X_g^i]}
 \label{eq:SW_map_for_X_0}
\end{align}
are related by a unitary transformation.  Here $\tau$ parametrizes the
trajectory $\theta(\tau)$, and $g(X,x)$ is assumed to be Hermitian for
any Hermitian $X$.

For $X_0$ and $X_g$ to be connected by a unitary transformation 
\begin{align*}
 X_g = u X_0  u^{-1},
\end{align*}
$u(\tau)$ must satisfy
\begin{align}
 \frac{{\rm d}u}{{\rm d}\tau}= iug(X_0,u^{-1} x u).
 \label{eq:condition_for_u}
\end{align}
Therefore the question is reduced to whether we can solve Eq.\
(\ref{eq:condition_for_u}) for a unitary operator $u(\tau)$.  

Let us expand as
\begin{align*}
 u(\tau)=e^{i\sum_{n=1}^\infty \tau^n H^{(n)}}, 
 \quad H^{(n)}{}^\dagger = H^{(n)}
\end{align*}
and try to determine $H^{(n)}$ order by order.  We also define
\begin{align*}
 u^{(m)}(\tau)=e^{i\sum_{n=1}^m \tau^n H^{(n)}}.
\end{align*}

Suppose that we have solved (\ref{eq:condition_for_u}) to
order $\tau^{m-2}$, namely, we have found a unitary $u^{(m-1)}$
satisfying
\begin{align*}
 \frac{{\rm d}u^{(m-1)}}{{\rm d}\tau}= iu^{(m-1)}g(X_0,{u^{(m-1)}}^{-1} x u^{(m-1)})
 + {\cal O}(\tau^{m-1}),
\end{align*}
or
\begin{align*}
 \left[ \frac{{\rm d}u^{(m-1)}}{{\rm d}\tau} \right]_{m-2}
 =  \left[ iu^{(m-1)}g(X_0,{u^{(m-1)}}^{-1} x u^{(m-1)}) \right]_{m-2}.
\end{align*}
Here we have defined
\begin{align*}
 \left[ \sum_{n=0}^\infty a_n \tau^n \right]_{m} \equiv \sum_{n=0}^m a_n \tau^n.
\end{align*}
Now try to solve (\ref{eq:condition_for_u}) at order $\tau^{m-1}$:
\begin{align*}
 \left[ \frac{{\rm d}u^{(m)}}{{\rm d}\tau} \right]_{m-1}
 =  \left[ iu^{(m)}g(X_0,{u^{(m)}}^{-1} x u^{(m)}) \right]_{m-1}.
\end{align*}
Since $[u^{(m)}]_{m}=[u^{(m-1)}]_m+i\tau^m H^{(m)}$, the left hand side
can be rewritten as
\begin{align*}
 \left[ \frac{{\rm d}u^{(m)}}{{\rm d}\tau} \right]_{m-1}
 &= \frac{{\rm d}}{{\rm d}\tau}[u^{(m)}]_m
 = \frac{{\rm d}}{{\rm d}\tau}[u^{(m-1)}]_m+im \epsilon^{m-1} H^{(m)}\\
 &= \left[ \frac{{\rm d}u^{(m-1)}}{{\rm d}\tau} \right]_{m-1} + im\tau^{m-1}H^{(m)}.
\end{align*}
Therefore
\begin{align*}
 im\tau^{m-1}H^{(m)}
 &= - \left[ \frac{{\rm d}u^{(m-1)}}{{\rm d}\tau} \right]_{m-1} 
 -  \left[ iu^{(m-1)}g(X_0,{u^{(m-1)}}^{-1} x u^{(m-1)}) \right]_{m-1},
\end{align*}
where we replaced $u^{(m)}$ with $u^{(m-1)}$ in $[~]_{m-1}$.  Now that
$H^{(m)}$ only appears on the left hand side, the question is whether
the right hand side is anti-Hermitian.  By assumption, the right hand
side is zero to order $\tau^{m-2}$.  Therefore, up to order $\tau^{m-1}$
we can multiply the right hand side by $[{u^{(m-1)}}^{-1}]_{m-1}$ to
obtain
\begin{align*}
 im\tau^{m-1}H^{(m)}
 &= - \left[ \frac{{\rm d}u^{(m-1)}}{{\rm d}\tau} {u^{(m-1)}}^{-1} \right]_{m-1} \\
 & \qquad -  \left[ iu^{(m-1)}g(X_0,{u^{(m-1)}}^{-1} x u^{(m-1)})
 {u^{(m-1)}}^{-1} \right]_{m-1} + {\cal O}(\tau^{m}).
\end{align*}
The two terms on the right hand side are easily shown to be anti-Hermitian
using the unitarity of $u^{(m-1)}$ and the assumption that $g(X,x)$ is
Hermitian for any Hermitian $X$.  Therefore $H^{(m)}$ is Hermitian.

Since $H^{(1)}=g$ is Hermitian, $H^{(n)}$ is Hermitian for all $n$ and
the proof is complete.

\section{Useful formulae}\label{section:useful_formulae}

Using relations
\begin{gather*}
 \Tr[e^{-\tau  \OP{N}} e^{-ik\cdot \OP{x}}]
 = \frac{e^{-\frac{\bar{k}k}{2}\coth\frac{\tau}{2}}}{(1-e^{-\tau})^d},\\
  \frac{\partial}{\partial k^{\alpha}} e^{-ik\cdot \OP{x}}
 =\left(-i\OP{a}^{\bar{\alpha}}-\frac{1}{2}k^{\bar{\alpha}}\right) e^{-ik\cdot \OP{x}},\quad
 \frac{\partial}{\partial k^{\bar{\alpha}}} e^{-ik\cdot \OP{x}}
 =\left(-i\OP{a}^{\alpha}+\frac{1}{2}k^{\alpha}\right) e^{-ik\cdot \OP{x}},
\end{gather*}
and $\frac{1}{\OP{N}+a}=\int_0^\infty {\rm d}\tau\,e^{-(\OP{N}+a)\tau}$, one can derive
\begin{align*}
 \Tr\left[\frac{1}{\hat{N}+a}e^{-ik\cdot \hat{x}}\right]
 &= 2^{1-d} \int_1^\infty \!\! {\rm d}y\, (y+1)^{d-a-1} (y-1)^{a-1}  e^{-\frac{\bar{k}k}{2} y},\\
 \int \! {\rm d}^{D}k\, e^{ik \cdot x} \,
 \Tr\left[\frac{1}{\hat{N}+a}e^{-ik\cdot \hat{x}}\right]
 &= 2(2\pi)^{d} \int_0^1 \!\! {\rm d}\eta\, (1+\eta)^{d-a-1} (1-\eta)^{a-1} 
 e^{-2\bar{z}z \eta},\\
 \Tr\left[\frac{1}{\hat{N}+a} \hat{a}^{\bar{\alpha}} \hat{a}^{\beta} e^{-ik\cdot \hat{x}}\right]
 &= 2^{-d} \Big[
 \delta^{\alpha\beta} \int_1^{\infty} \!\! {\rm d}y\, (y+1)^{d-a-1} (y-1)^{a} e^{-\frac{\bar{k}k}{2} y}\\
 & \qquad\qquad - \frac{k^{\bar{\alpha}}k^{\beta}}{2}
 \int_1^{\infty} \!\! {\rm d}y\, (y+1)^{d-a} (y-1)^{a} e^{-\frac{\bar{k}k}{2} y}
 \Big],\\
 \int\!  {\rm d}^{D}k\, e^{ik \cdot x} \,
 \Tr\left[\frac{1}{\hat{N}+a} \hat{a}^{\bar{\alpha}} \hat{a}^{\beta} e^{-ik\cdot \hat{x}}\right]
 &= (2\pi)^d \Big[
 - \delta^{\alpha\beta} \int_0^1 \!\! {\rm d}\eta\,
 (1+\eta)^{d-a-1} (1-\eta)^a   e^{-2 \bar{z}z \eta}\\
 & \quad\qquad\qquad + 2 x^{\bar{\alpha}} x^{\beta}
 \int_0^1 \!\! {\rm d}\eta\,
 (1+\eta)^{d-a} (1-\eta)^a   e^{- 2 \bar{z}z \eta} \Big],
\end{align*}
where $k\cdot\OP{x}\equiv k^{\bar{\alpha}} \OP{a}^{\alpha}+k^{\alpha}
\OP{a}^{\bar{\alpha}}$, $k\cdot x\equiv k^{\bar{\alpha}} z^{\alpha}+k^{\alpha}
\bar{z}^{\bar{\alpha}}$, $\bar{k}k\equiv k^{\bar{\alpha}}k^{\alpha}$, and
$\bar{z}z\equiv \bar{z}^{\bar{\alpha}}z^{\alpha}$.

\end{document}